\documentstyle[12pt,frascatiphys1]{article}
\begin{document}
\title{ ON  RADIATIVE $\phi \to \eta \gamma$, $\phi \to \eta^\prime
\gamma$ DECAYS\thanks{~Presented by F. De Fazio}}
\author{Fulvia De Fazio \\
{\em Centre for Particle Theory, University of Durham, Durham DH1 3LE, 
U.K.} \\
M.R.~Pennington        \\
{\em Centre for Particle Theory, University of Durham, Durham DH1 3LE, 
U.K.} }
\maketitle
\baselineskip=14.5pt
\begin{abstract}
We  use QCD sum-rules to study the decays
$\phi \to \eta \gamma$ and $\phi \to \eta^\prime \gamma$, obtaining
${\cal B}( \phi  \to \eta \gamma)=(1.15 \pm 0.2)~10^{-2}$ and
${\cal B}( \phi \to \eta^\prime \gamma)=(1.18 \pm 0.4)~
10^{-4}$, in very good agreement
with existing experimental data.
We also discuss  the issue of  $\eta-\eta^\prime$ mixing, 
predicting the $\eta$ and $\eta^\prime$ decay constants in 
 a  mixing scheme in the quark-flavour basis.
\end{abstract}
\baselineskip=17pt
\section{Introduction}
The reasons for interest in  $\phi$ radiative
decays are manifold. These decays yield  
information on low energy hadron physics, 
providing insight into the  controversial issue of
$\eta-\eta^\prime$ mixing. From the experimental side,  
new data
are expected from the KLOE experiment at the DA$\Phi$NE
$\phi$-factory\cite{book}, where a large sample of $\phi$ decays will be
collected,
 which will  considerably improve the presently available 
statistics\cite{novos}. 

In the following 
we study $\phi \to \eta \gamma$,  $\phi \to \eta^\prime \gamma$ decays
using QCD sum-rules\cite{shifman}.
After surveying  $\eta-\eta'$ mixing,
 we consider  the coupling of the strange pseudoscalar current
 to  $\eta$ and $\eta^\prime$, preliminary to our
analysis of $\phi$ decays. Though we 
work with  interpolating currents defined in the quark flavour basis,
our results do not depend on any  mixing scheme, providing
mixing scheme independent QCD predictions.\\
We also consider  the
coupling of $\eta$ and $\eta^\prime$  to the strange and non-strange axial
currents, identifying the results with the decay constants in the flavour
basis mixing scheme and estimating the
mixing parameters.
Finally, we give conclusions.
\section{On $\eta-\eta^\prime$ Mixing}
 $\eta-\eta^\prime$ mixing is
a much debated subject. The once conventional description introduced
a single mixing
angle in the octet-singlet flavour basis\cite{feldrev}. 
More recently, it has been shown\cite{leutproc}
that a proper treatment 
requires  two  angles with a    
redefinition of the particle decay constants. An equivalent description
adopts 
the quark-flavour basis instead of the octet-singlet 
one\cite{feld}, in which such constants are defined as: 
\begin{equation}
<0|J_{5 \mu }^a|P(p)>=i ~f_P^a p_\mu \;\;\; (a=q,s; ~~~
P=\eta,\eta^\prime) \;,
\label{lepconst} \end{equation}
\noindent with 
$J_{5 \mu }^q = {1 \over \sqrt{2}}({\overline u} \gamma_\mu \gamma_5 u+   
{\overline d} \gamma_\mu \gamma_5 d)$,
$J_{5 \mu }^s= {\overline s} \gamma_\mu \gamma_5 s$.
The decay constants are
written  according to  the following mixing pattern:
\begin{eqnarray}
f_\eta^q=f_q ~\cos \phi_q && \hskip 1 cm f_\eta^s=-f_s ~ \sin \phi_s
\nonumber \\
f_{\eta^\prime}^q=f_q ~\sin \phi_q && \hskip 1 cm f_{\eta^\prime}^s=f_s ~
\cos \phi_s \;\;\;\;.
\label{constmix}
\end{eqnarray}
\noindent 
Feldmann\cite{feld} has shown that, since
$|\phi_s-\phi_q|/ (\phi_s+\phi_q)\ll 1$, this mixing framework
is simpler, being
specified
quite accurately in terms of a
single  angle $\phi=\phi_q=\phi_s$.

\section{$\eta$ and $\eta^\prime$ couplings to the pseudoscalar
current}
Let us define:
$ <0|{\overline s} i\gamma_5 s|\eta>=A $,
and compute this
quantity using QCD sum-rules starting from the two-point correlator:
\begin{equation}
T_A(q^2)=i ~\int d^4 x e^{iq \cdot x} <0|T[ J_{5}^s(x) J_{5}^{s
\dagger}(0)]|0>
\label{cor_a}
\end{equation}
\noindent
where $J_5^s={\overline s} i\gamma_5 s$. The correlator (\ref{cor_a})
is given by the dispersive representation:
\begin{equation}
T_A(q^2)={1 \over \pi} \int_{4 m_s^2}^\infty \, ds\;{\rho(s) \over s-q^2}
+{\rm subtractions} \;. 
\label{disp}
\end{equation}
\noindent For low values of $s$, $\rho(s)$  receives contribution from the
coupling of
the $\eta$ to the pseudoscalar current. Hence we
can write (we discuss later possible subtractions):
\begin{equation}
T_A(q^2)={ A^2 \over m_\eta^2 -q^2}+ {1 \over \pi} \int_{s_0}^\infty\,ds\;
{\rho^{had}(s) \over s-q^2} \;,
\end{equation}
\noindent assuming that higher
 contributions  start from an effective threshold
$s_0$.
$T_A(q^2)$  can also be computed in QCD by
expanding the $T$-product in (\ref{cor_a}) by an Operator Product
Expansion as the sum of a perturbative term plus 
non-perturbative ones  proportional to vacuum 
condensates. 
Dispersively writing  
the perturbative term too, we have:
\begin{equation}
T_A^{QCD}(q^2)={1 \over \pi} \int_{4 m_s^2}^\infty\,ds\; {\rho^{QCD}(s) \over
s-q^2} + d_3 <{\overline s } s> +d_5<{\overline s}g \sigma  G s>+...
\label{qcd_a} \; .
\end{equation}
\noindent   $\rho^{QCD}$ and 
$d_3$, $d_5$ are computed in QCD.
Now we  implement quark-hadron duality,
assuming that $\rho^{had}(s)$ and $\rho^{QCD}(s)$ 
give the same result 
when integrated above  $s_0$. This leads to the sum-rule:
\begin{equation}
{ A^2 \over m_\eta^2 -q^2}={1 \over \pi} \int_{4 m_s^2}^{s_0}\,ds\; 
{\rho^{QCD}(s) \over s-q^2} + d_3 <{\overline s } s> +d_5<{\overline s}g \sigma G s>+...
\label{sr} 
\end{equation}
\noindent  Now we apply to both sides of (\ref{sr})
a Borel transform, defined as 
\begin{equation}
{\cal B} [f(Q^2)]=lim_{Q^2 \to \infty, \; n \to \infty, \; {Q^2 \over
n}=M^2}\;
{1 \over (n-1)!} (-Q^2)^n \left({d \over dQ^2} \right)^n f(Q^2) \; ,
\label{tborel}
\end{equation}
where $f$ is a generic function of $Q^2=-q^2$ and  
 $M^2$ is
known as the Borel parameter. This operation
improves the
convergence of the sum rule and eliminates the contribution 
of subtraction terms in (\ref{disp}).
The final sum-rule reads:
\begin{eqnarray}
&& A^2 e^{-{m_\eta^2 \over M^2}}= {3 \over 8 \pi^2} \int_{4 m_s^2}^{s_0}
ds~s ~\sqrt{1-{4 m_s^2 \over s}}e^{-{s \over M^2}}
\nonumber \\
&-& m_s  e^{-{m_s^2 \over M^2}} \Bigg[ <{\overline s} s>\Bigg( 1-{m_s^2 \over
M^2}+
{m_s^4 \over M^4} \Bigg)+ {1 \over M^2} <{\overline s}g \sigma G s> 
\Bigg( 1-{ m_s^2 \over 2 M^2} \Bigg) \Bigg] 
\label{finsr} \; .
\end{eqnarray}
\noindent 
We use $<{\overline s}g \sigma G
s>=0.8~$GeV$^2 <{\overline s}s>$, $<{\overline s}s>=0.8 <{\overline
q}q>$,  
$<{\overline q}q>=(-0.24)^3$ GeV$^3$, 
$m_\eta=0.548$ GeV. We vary the strange quark mass in the range:
 $m_s=0.125-0.140$ GeV\cite{noistrano} and $s_0$
below the $\eta^\prime$ pole between 
$0.9^2-0.95^2$ GeV$^2$. 
Since $M^2$ is an  unphysical parameter, we look
for  a range of its values  (\lq\lq stability
window") where the  
sum-rule is almost independent of $M^2$. 
We fix 
$M^2$ in $[0.8,1]$ GeV$^2$ and, 
considering  the uncertainty on $m_s$ and on $s_0$, we
get\cite{noi}:
\begin{equation}
|A|=(0.115 \pm 0.004)\; {\rm GeV}^2 \label{a} \; .
\end{equation}
Let us now consider:
$<0|{\overline s} i\gamma_5 s|\eta^\prime>=A^\prime$.
An analogous calculation gives:
\begin{eqnarray}
&& (A^\prime)^2e^{-{m_{\eta^\prime}^2 \over M^2}} + A^2 e^{-{m_\eta^2
\over M^2}}= {3 \over 8 \pi^2} \int_{4 m_s^2}^{s_0^\prime} ds
~s ~\sqrt{1-{4 m_s^2 \over s}}e^{-{s \over M^2}}
\nonumber \\
&-& m_s  e^{-{m_s^2 \over M^2}} \Bigg[
<{\overline s} s>\Bigg( 1-{m_s^2 \over M^2}+
{m_s^4 \over M^4} \Bigg)+ {1 \over M^2}<{\overline s}g \sigma G s> 
\Bigg( 1-{ m_s^2 \over 2 M^2} \Bigg) \Bigg] \; ,
\end{eqnarray}
\noindent where we have raised the  threshold up to $s_0^\prime$
 as to pick up the $\eta^\prime$ pole too.
Using $m_{\eta^\prime}=0.958$ GeV and in
the stability window  $[1.2,2]$ GeV$^2$ for $M^2$, 
we obtain\cite{noi}: 
\begin{equation}
|A^\prime|=(0.151 \pm 0.015)\; {\rm GeV}^2 \; .
\label{ap}
\end{equation}
\noindent 
Though we cannot actually establish the sign of $A$,
$A^\prime$, we assume that $A \cdot A^\prime >0$.

\section{Radiative $\phi \to \eta \gamma$ and 
 $\phi \to \eta^\prime \gamma$ decays}

Let us define:
$<\eta(q_2)|{\overline s} \gamma^\nu s|\phi(q_1,\epsilon_1)>=F(q^2)~
\epsilon^{\nu \alpha \beta \delta} (q_1)_\alpha (q_2)_\beta
(\epsilon_1)_\delta$,
($q=q_1-q_2$).
In order  to compute  $\phi \to \eta \gamma$  decay, we need the
coupling $g=-{1 \over 3} F(0)$, obtained for a real photon coupling to a
strange quark.
We  consider:
\begin{equation}
\Pi_{\mu \nu}=i^2 \int d^4 x~ d^4 y~ e^{-i q_1 \cdot
x+i q_2 \cdot y} <0|T[J_5^s(y) J_\nu(0) J_\mu(x)]|0> 
= 
\Pi(q_1^2,q_2^2,q^2)\, 
\epsilon_{\mu \nu \alpha \beta} 
q_1^\alpha q_2^\beta\label{cor-phi}
\end{equation}
with $J_5^s$  defined above and $J_\nu={\overline s} \gamma_\nu s$. 
The  sum-rule is built up for 
$\Pi(q_1^2,q_2^2,q^2)$.
First we   write $\Pi(q_1^2,q_2^2,q^2)$ according to a dispersion
relation in
the variables $q_1^2,q_2^2$:
\begin{equation}
 \Pi(q_1^2,q_2^2,q^2)={1 \over \pi^2} \int ds_1 \int ds_2
{\rho(s_1,s_2,q^2) \over (s_1 -q_1^2)(s_2-q_2^2)} \; .
\end{equation}
Now the spectral function
contains,
for low values of $s_1$,
$s_2$, a double $\delta-$function corresponding to the transition $\phi \to
\eta$. Extracting this contribution, 
we derive the
sum-rule (after a double Borel transform in the variables $-q_1^2$,
 $-q_2^2$):
\begin{eqnarray}
A\; F(q^2) m_{\phi} f_{\phi} &=&
e^{{m_{\phi}^2 \over M_1^2}} e^{{m_{\eta}^2 \over M_2^2}}
 \Bigg\{ \int d s_1 \int d s_2  
e^{-{s_1 \over M_1^2}} e^{-{s_2 \over M_2^2}}
 { 3 m_s \over \pi^2 \sqrt{\lambda(s_1,s_2,q^2)} }
\nonumber \\
+ e^{-{m_s^2\over M_1^2}}e^{-{m_s^2 \over M_2^2}}
 \Bigg[ <{\overline s} s> && \left(2 -{m_s^2 \over M_1^2}-{m_s^2 \over M_2^2}
+{m_s^4 \over M_1^4}+{m_s^4 \over M_2^4}+{m_s^2(2m_s^2-q^2)\over M_1^2
M_2^2} \right) \nonumber \\
+ <{\overline s } g \sigma G s>&& \left( {1 \over 6 M_1^2} +{2 \over 3 M_2^2}
-
{ m_s^2 \over 2 M_1^4}-{m_s^2 \over 2 M_2^4} +{(2 q^2 -3 m_s^2)
\over 3 M_1^2 M_2^2} \right) \Bigg] \Bigg\} \label{srb}
\end{eqnarray}
The integration domain over  $s_1,s_2$ 
(specified in\cite{noi})  depends on $q^2$.
We compute  $F(q^2)$ for negative
values of $q^2$, and 
then  extrapolate the result to $q^2=0$ 
\footnote{In this way, we could perform a double Borel transform in the
two
variables $Q_1^2=-q_1^2$ and $Q_2^2=-q_2^2$, which allows us to remove  
single poles in the $s_1$ and $s_2$ channels
(``parasitic'' terms).}. 
Since we
only know the magnitude of $A$, it is 
$|F(q^2)|$ that is determined.
We use:
$ m_\phi=1.02$ GeV,
$f_\phi=0.234$ GeV (obtained from the experimental datum on
$\phi \to e^+
e^-$\cite{pdg}).
We use two values of
the $\phi$ threshold: $s_{01}=1.8,1.9$ GeV$^2$;
the $\eta$ threshold is chosen as 
in section~3.
The extrapolation to $q^2=0$ gives
$|g|=F(0)/3=(0.66 \pm 0.06)\,{\rm GeV}^{-1}$;
 the uncertainty is obtained  varying all
the input parameters  in the sum rule. 
From this result we compute  $\Gamma(\phi \to \eta \gamma)$ 
and, using $\Gamma(\phi)=4.43$ MeV\cite{pdg},  find\cite{noi}:
\begin{equation}
{\cal B}(\phi \to \eta \gamma)  =(1.15 \pm 0.2)\% \; ,
\label{breta}
\end{equation}
\noindent in agreement with 
the experimental result:
${\cal B}(\phi \to \eta \gamma) = (1.18 \pm 0.03 \pm 0.06) \%$\cite{novos}.
Applying the same analysis to the $\eta^\prime$ mode gives
$|g^\prime|=F^\prime(0)/3=(1.0 \pm 0.2)\; {\rm GeV}^{-1}$ and\cite{noi}
\begin{equation}
{\cal B}(\phi \to \eta^\prime \gamma) = (1.18 \pm 0.4) ~10^{-4}
\label{bretap} \; ,
\end{equation}
\noindent
which agrees with the experimental datum  
${\cal B}(\phi \to \eta^\prime \gamma) = (0.82^{+0.21}_{-0.19} \pm
0.11)~10^{-4}$\cite{novos}.
The results  (\ref{breta}) and
(\ref{bretap})  are  independent 
of any mixing scheme for the $\eta$ and $\eta^{\prime}$.
Nevertheless, adopting the mixing scheme in the flavour basis, 
one gets:
\begin{equation}
R={ {\cal B}(\phi \to \eta \gamma)  \over  {\cal B}(\phi \to
\eta^\prime 
\gamma)} =\Bigg({m_\phi^2-m_\eta^2 \over m_\phi^2-m_{\eta^\prime}^2}
\Bigg)^3 \tan^2 \phi_s 
\end{equation}
and hence $\phi_s=(34 \pm^8_6 )^o$.
The experimental ratio would give: $\phi_s=(39.0 \pm^{7.5}_{5.5})^o$.
For a comprehensive survey of other results, we refer
 to\cite{feldrev}.

\section{$\eta$ and $\eta^\prime$ couplings to  axial currents}

Now we  
investigate the $\eta$, $\eta^\prime$ decay constants
in both the strange and non-strange sectors 
in order to deduce the mixing pattern. 
  We begin by considering
the correlator
\begin{equation}
\Pi_{\mu \nu}(q^2)=i ~\int d^4 x~ e^{iq \cdot x} <0|T[ J_{5
\mu}^s(x) J_{5 \nu}^s(0)]|0> \;\; .
\end{equation}
Following the procedure already outlined above, we obtain the
sum-rule:
\begin{equation}
(f_\eta^s)^2=e^{m_\eta^2 \over M^2}  \Bigg[ {1 \over 4 \pi^2} \int_{4
m_s^2}^{s_0}
ds ~e^{-{s \over M^2}} 
\sqrt{1 -{4 m_s^2 \over s}}\quad {2 m_s^2+s
\over s}
+{2 m_s \over M^2} <{\overline
s}s> e^{-{m_s^2 \over M^2}}
\Bigg]\; .
\end{equation}
In the  stability window  $[2,3.5]$ GeV$^2$ for $M^2$ 
we get\cite{noi}
$f_\eta^s=(0.13 \pm 0.01)\; {\rm GeV}$ and, using 
the same technique:
$f_{\eta^\prime}^s=(0.12 \pm 0.02)\; {\rm GeV}$,
 $f_\eta^q=(0.144 \pm 0.004)\;{\rm GeV}$,
$f_{\eta^\prime}^q=(0.125 \pm 0.015)\;{\rm GeV}$.
From these  results we can 
estimate all the mixing parameters defined 
in (\ref{constmix}), obtaining:
 $\phi_s=(46.6^o\pm 7^o)$, $f_s=(0.178 \pm 0.004)$~GeV
and
 $\phi_q=(41^0 \pm4^o)$, $f_q=(0.19 \pm 0.015)$~GeV.
Our results give  
$|\phi_s -\phi_q|/(\phi_s +\phi_q) \simeq 0.065$, 
 confirming 
  that this ratio is much less than 1\cite{feld}.

Finally, let us consider  the matrix element:
$<0|\partial^\mu J_{5 \mu }^s|\eta>=m_\eta^{\,2} f_\eta^s$.
 The divergence of the axial current contains the 
anomaly:
$\partial^\mu J_{5 \mu }^s=\partial^\mu ({\overline s} \gamma_\mu
\gamma_5 s)=
2 m_s {\overline s} i\gamma_5 s +{\alpha_s \over 4 \pi} G \tilde G$,
 $G$ being the gluon field strength tensor,  ${\tilde G}$ 
its dual. Therefore:
\begin{equation}
2 m_s <0|{\overline s} i\gamma_5 s| \eta^{(\prime)}>=f_{\eta^{(\prime)}}^s
m^2_{\eta^{(\prime)}}-\langle 0 \left|{\alpha_s \over 4 \pi}G \tilde
G\right|\eta^{(\prime)} \rangle
\;. \label{glue}
\end{equation}
 \noindent Exploiting  the  previous results 
and (\ref{a}), (\ref{ap}), we derive from
(\ref{glue}):
\begin{equation}
\langle 0 \left|{\alpha_s \over 4 \pi}G \tilde G\right
|\eta\rangle =(0.008 \pm 0.004)\,{\rm GeV}^3 \;\; ,
\langle 0 \left|{\alpha_s \over 4 \pi}G \tilde G
\right|\eta^{\prime} \rangle =(0.072 \pm 0.025)\,{\rm GeV}^3
\label{gg} .
\end{equation}
\noindent
Both values in (\ref{gg}) are close to the  quark model 
result of Novikov et al.\cite{novikov}.

\section{Conclusions}

We have analysed  $\phi \to \eta \gamma$, $\phi \to \eta^\prime
\gamma$ decays using QCD sum-rules. We begin by presenting
a preliminary
calculation of the couplings of the pseudoscalar current to  $\eta$, 
$\eta^\prime$. The results  derived require  no assumption 
about $\eta-\eta^\prime$ mixing and 
  are in good agreement with 
available data.
However, the 
uncertainty in the $\eta^\prime$ case is large, and so the last word is
left to the experimental
improvement  at DA$\Phi$NE, for instance. 

We have also considered  $\eta-\eta^\prime$
mixing, estimating 
the mixing parameters in a
quark-flavour basis  scheme. The existing
spread of results gives us
confidence that new experimental information will shed light on this
topic too.

We are most grateful for support from the EU-TMR Programme, 
Contract No.
CT98-0169, EuroDA$\Phi$NE.

\end{document}